\documentclass[twocolumn,showpacs,preprintnumbers,amsmath,amssymb,aps,prl]{revtex4}
\usepackage{graphicx,graphics}
\usepackage{dcolumn}
\usepackage{bm}
\usepackage{psfrag}

\begin{document}
\title{Mott-insulator phases of spin-3/2 fermions\\
in the presence of quadratic Zeeman coupling}
\author{K. Rodr\'\i guez}
\author{A. Arg\"uelles}
\author{M. Colom\'e-Tatch\'e}
\author{T. Vekua}
\author{L. Santos}
\affiliation{Institut f\"ur Theoretische Physik, Leibniz Universit\"at Hannover, Appelstr. 2 D-30167, Hannover, Germany}
\begin{abstract}
We study the influence of the quadratic Zeeman effect in the Mott-insulator phases of hard-core 
spin-$3/2$ fermions. We show that contrary to spinor bosons, any quadratic Zeeman coupling 
preserves a $SU(2)\otimes SU(2)$ symmetry,  
leading for large-enough quadratic Zeeman coupling to an isotropic pseudo-spin-1/2 Heisenberg antiferromagnet. 
Depending on the scattering lengths, on 1D lattices the quadratic Zeeman 
coupling can induce either a Kosterlitz-Thouless transition between a gapped dimerized spin-$3/2$ 
phase and a gapless pseudo-spin-1/2 antiferromagnet, or a commensurate-incommensurate transition 
from a gapless spin-liquid into the pseudo-spin-1/2 antiferromagnet. 
Similar arguments allow to foresee corresponding transitions on ladder type and square lattices.
We analyze various observables which should reveal in experiments these phases. 
\end{abstract}


\maketitle



Ultra-cold gases in optical lattices constitute an extraordinary tool 
for the analysis of strongly-correlated gases under 
extremely well-controlled conditions~\cite{Lewenstein2006,Bloch2008}, 
as highlighted by the observation of the superfluid to Mott-insulator (MI) transition 
in ultra-cold bosons~\cite{Greiner2002}, recently followed by the realization 
of the metal to MI transition in two-component fermions~\cite{Jordens2008,Schneider2008}. 
Due to super-exchange, the MI phase of spin-$1/2$ fermions 
is expected to acquire a magnetic Ne\'el (antiferromagnetic) ordering, whose observation is the 
goal of active on-going efforts~\cite{Jordens2009}.

Optical traps permit the simultaneous trapping of various Zeeman sublevels, allowing for 
multi-component (spinor) gases. Spinor bosons have attracted a large interest due to their 
rich ground-state physics and spinor dynamics~\cite{Ho1998}.
Recently, spinor fermions are also attracting a rapidly-growing attention, motivated by
experiments on BEC-BCS crossover in two-component fermions~\cite{Bloch2008} and the availability of 
multi-component fermions, including three-component Li gases~\cite{Wenz2009}, four-component $^{40}$K~\cite{DeMarco2001}, spin-$3/2$ fermions as $^{135}$Ba 
and $^{137}$Ba~\cite{He1991}, and Fermi-degenerate Yb~\cite{Fukuhara2007}. Multi-component fermions present a wealth of novel phases. Pseudo-spin-$1$ 
fermions~\cite{Honerkamp2004} 
allow for color superfluidity and trions, whereas spin-$3/2$ gases 
are even richer~\cite{Wu2003,Lecheminant2005,Wu2005,Controzzi2006}, 
including quartets. Ultra-cold Yb opens the fascinating possibility 
of $SU(6)$-symmetric spin-$5/2$ gases~\cite{Cazalilla2009}.

A rich physics is also expected for repulsive spin-$3/2$ fermions~\cite{Lecheminant2005,Wu2005}, 
which in quarter filling may undergo a 
MI transition. Whereas for spin-$1/2$ a Ne\'el phase is expected,  
the MI in spin-$3/2$ presents a richer magnetic structure given by 
a gapless spin-liquid or a gapped dimerized phase, depending on 
the interatomic interactions. 
Contrary to the spin-$1/2$ case, where spin-changing collisions are absent and the 
quadratic Zeeman effect (QZE) is irrelevant, the QZE is crucial
for higher-spins, as shown in spinor condensates~\cite{Ho1998}. 
However, in spite of its experimental relevance, the QZE has been typically 
ignored in the analysis of the magnetic properties. 

\begin{figure}
\vspace*{-0.5cm}
\includegraphics[width=8.0cm]{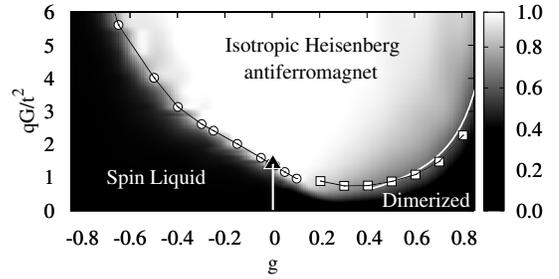}\\
\vspace*{-0.5cm}
\caption{Phase diagram for spin-$3/2$ fermions with QZE. We depict the 
chirality $\tau$ (density plot), 
the singlet-triplet crossing in the excitation spectrum (squares), 
the jump of the exponent $\gamma$ to $-1$ (circles),   
the critical $q_c$ expected from two-band theory (triangle), and the $J_2/J=0.25$ line 
resulting from a strong-coupling analysis (white curve). See text.}
\vspace*{-0.3cm}
\label{fig:1}
\end{figure}

In this Letter, we explore, for the 
first time to our knowledge, the role of the QZE in the magnetism of spin-$3/2$ MI phases. 
We show that contrary to spinor bosons~\cite{Kolezhuk2008}, any QZE preserves a $SU(2)\otimes SU(2)$ 
symmetry which results at large QZE in a pseudo-spin-$1/2$ isotropic Heisenberg antiferromagnet (iHAF).  
Depending on the scattering lengths, our results suggest for growing QZE
a Kosterlitz-Thouless (KT) phase transition (PT) from a gapped dimerized spin-$3/2$
phase into the iHAF, or a commensurate-incommensurate (C-IC) PT between a gapless spin-liquid again into the iHAF.
We analyze in detail observables 
which may characterize the phases (Fig.~\ref{fig:1}) and PTs in future experiments~\cite{Sitte2009}.

We consider a balanced mixture of spin-$3/2$ fermions in a 1D lattice~\cite{footnote-1}
with $N_m=N_{-m}$ ($N_m=$ number of fermions with spin projection $m$), such that the magnetization 
${\cal M}\equiv\sum_m m N_m=0$. 
For a deep lattice and low filling (single-band regime) 
the system Hamiltonian is 
\begin{eqnarray}
\hat H \!\!&=&\!\! -t\sum_{m,i}\left [ \hat\psi_{m,j}^\dag\hat\psi_{m,j+1} +{\mathrm H.c.} \right ] -q\sum_{m,j} m^2\hat n_{m,j}\nonumber \\
&+& g_0 \sum_i P_{00,i}^\dag P_{00,i}+g_2\sum_{i,m_F}P_{2m_F,i}^\dag P_{2m_F,i},
\label{eq:H}
\end{eqnarray} 
where $\hat \psi_{m,j}$ annihilate fermions with spin $m$ at site $j$, 
$P_{Fm_F,i}^\dag=\sum_{m_1,m_2} \langle F,m_F|m_1,m_2\rangle \psi_{m_1,i}^\dag\psi_{m_2,i}^\dag$, with 
$\langle F,m_F|m_1,m_2\rangle$ the Clebsch-Gordan coefficients, and $\hat n_{m,j}=\psi_{m,j}^\dag\psi_{m,j}$. 
$t$ is the hopping rate between adjacent sites, and  
$g_{0,2}=4\pi\hbar^2a_{0,2}/M$ characterize the $s$-wave channels 
with total spin $0$ and $2$ (the only available due to symmetry), with $a_{0,2}$
the scattering lengths and $M$ the atomic mass. Although typically $a_0$ 
and $a_2$ are similar, their values may be variated by means of 
microwave dressing~\cite{Papoular2009} or optical Feshbach resonances~\cite{Fedichev1996}. 
Below we use $G=(g_0+g_2)/2$, and $g=(g_2-g_0)/(g_2+g_0)$. 
The interactions preserve ${\cal M}$ and hence the linear 
Zeeman effect induced by a magnetic field
does not play any role for a fixed ${\cal M}$.
However spin-changing collisions (which re-distribute the populations of the different components while preserving ${\cal M}$) are crucial. Finally, the QZE, characterized by the externally controllable (by means of a magnetic field or microwave or optical dressing)
constant $q$, is crucial in spinor gases, and 
plays a key role below. 

For large-enough interactions, $G\gg t$, we may 
consider the hard-core case, with maximally one fermion per site. For a chemical potential 
$\mu$ larger than a critical $\mu_c(t)$ 
the system enters into the MI regime with one fermion per site. We 
focus below in the magnetic properties of these MI phases.
Within the MI regime we perform perturbation theory 
obtaining an effective Hubbard Hamiltonian 
with super-exchange interactions, which for two adjacent sites $\{1,2\}$ is of the form 
$\hat H_{12}=\hat H_0+\hat H_{sc}$, where
\begin{eqnarray}
&&\!\!\!\hat H_0=-\sum_{m} q m^2(\hat n_{m,1}+\hat n_{m,2}) \nonumber \\
&&\!\!+c_2\!\!\sum_{|m| \neq |m'|}\left (
\hat n_{m,1}\hat n_{m',2}-\hat \psi_{m,1}^\dag\hat \psi_{m',2}^\dag\hat \psi_{m,2}\hat \psi_{m',1}
\right ) \nonumber \\
&&\!\!\!+\!\!\!\sum_{|m|=1/2}^{3/2}\!\!c_{|m|}\!\left (
\hat n_{m,1}\hat n_{-m,2}-\hat \psi_{m,1}^\dag\hat \psi_{-m,2}^\dag\hat \psi_{m,2}\hat \psi_{-m,1}
\right )
\end{eqnarray}
contains self-energies and effective hoppings 
without spin-changing, whereas a simultaneous hopping and spin-changing is given by
$\hat H_{sc}=c_{sc}\left (\hat B_{1/2}^\dag\hat B_{3/2}+{\mathrm{H.c.}} \right )$,
with $\hat B_m=\hat\psi_{m,2}\hat\psi_{-m,1}-\hat\psi_{-m,2}\hat\psi_{-m,1}$. 
We introduce above:
\begin{eqnarray}
&&c_{3/2}\equiv -2t^2 \left (
\frac{\cos^2\phi}{9q/2+\lambda_+}+\frac{\sin^2\phi}{9q/2+\lambda_-}
\right ), \\
&&c_{1/2}\equiv -2t^2 \left (
\frac{\cos^2\phi}{q/2+\lambda_-}+\frac{\sin^2\phi}{q/2+\lambda_+}
\right ), \\
&&c_{\mathrm{sc}}\equiv t^2 \sin 2\phi \sum_{\beta=\pm} \eta_\beta
\frac{5q/2+\lambda_\beta}{\prod_{{\bar m}=1/2}^{3/2}(2q{\bar m}^2+\lambda_\beta)},
\end{eqnarray}
with $\eta_\pm=\pm 1$, 
$\lambda_\pm\equiv G-5q/2\pm \left [ 4q^2+g^2G^2 \right ]^{1/2}$, and 
$\tan\phi=\left([4q^2+g^2G^2]^{1/2}+2q\right )/gG$. 

Due to the spin-changing collisions, the single-site 
two-particle eigenstates with zero magnetization 
are not states with spins $m$ and $-m$ ($|m,-m\rangle$), but 
the linear combinations $|+\rangle=\cos\phi|-3/2,3/2\rangle+\sin\phi|-1/2,1/2\rangle$ and 
$|-\rangle=-\sin\phi|-3/2,3/2\rangle+\cos\phi|-1/2,1/2\rangle$ (with respective eigenenergies 
$\lambda_\pm$). This mixing leads to $\hat H_{sc}$ and 
the non-trivial $q$ and $g$ dependence of $c_{3/2}$, $c_{1/2}$, and $c_{sc}$. 
Note that $c_{3/2}$, $c_{1/2}$, $c_{sc}$ may diverge 
at specific values for which spin-changing collisions and 
QZE enter in resonance. Below we consider only non-resonant situations 
for which the hard-core formalism remains valid.



For $q=0$ the ground-state of $H=\sum_i H_{i,i+1}$ only depends on $g$~\cite{Lecheminant2005,Wu2005}. 
For $-1<g\leq 0$ the ground-state is a gapless spin liquid phase (with $3$ gapless spin-modes). 
This phase includes the exactly solvable $SU(4)$ line ($g=0$, i.e. $g_0=g_2$)~\cite{Sutherland1975}.
For $0<g<1$ the ground-state is a dimerized (spin-Peierls) phase which 
exhibits a spin gap and finite long-range  
dimer-dimer correlations $\lim_{n\rightarrow\infty}\langle D_iD_{i+n} \rangle = f_0$, with 
$\hat D_i=(-1)^i\hat{\bf S}_i(\hat{\bf S}_{i-1}+\hat{\bf S}_{i+1})$, with $\hat {\mathbf S}$ the 
spin-$3/2$ operators.  
At $g=0$ the system undergoes a KT-like transition between both phases.



The $\pm 1/2$ and $\pm 3/2$ manifolds 
are just linked by $\hat H_{sc}$. Hence, at large-enough $q>0$, when 
$\epsilon\equiv c_{sc}/(c_{1/2}-c_{3/2}-2q)\ll 1$, the $\pm 3/2$ manifold is favored, 
and the system acquires a pseudo-spin-$1/2$ character. Below we restrict to $q>0$, 
but similar reasonings apply to $q<0$, for which the $\pm 1/2$ manifold is favored. 
Projecting onto the $\pm 3/2$ manifold and 
up to second order in $\epsilon>0$, the Hamiltonian reduces  
to an isotropic Heisenberg form, $\hat H_{12}\simeq J \hat{\bf S}_1\cdot\hat{\bf S}_2$,
where $J\equiv -2(c_{3/2}-2\epsilon c_{sc})$ and ${\bf S}$ denotes pseudo-spin-$1/2$ operators 
$\hat S_z\equiv (\hat n_{3/2}-\hat n_{-3/2})/2$, $\hat S_+\equiv \hat\psi_{3/2}^\dag\hat\psi_{-3/2}$ and 
$\hat S_-\equiv\hat S_+^\dag$. Since $J>0$ for all $g_0,g_2,q>0$, 
the system reduces to an iHAF for large-enough $q$. This is true at any order in $\epsilon$ due to a hidden $SU(2)$ 
symmetry (see below). This must be compared to the case of spinor bosons~\cite{Kolezhuk2008}, 
where virtual transitions between manifolds lead for large $q$ to an anisotropic Heisenberg Hamiltonian, 
demanding a fine tuning of the microscopic constants to map the system to an iHAF.

Interestingly, spin-$3/2$ fermions, which show a hidden $SO(5)$ symmetry for $q=0$~\cite{Wu2003}, retain at any $q$ a $SU(2)\otimes SU(2)$ symmetry, generated by a direct product of two $SU(2)$ spin algebras operating in the $\pm 1/2 $ and $\pm 3/2$ manifolds, respectively. The $SU(2)$ generators are 
$\hat S^{\alpha}_z= 1/2\sum_i( \hat n_ {\alpha,i}- \hat n_ {-\alpha,i} )$, 
$ \hat S^{\alpha}_+= \sum_i \psi_{\alpha,i}^{\dagger}\psi_{-\alpha,i} $, with $\alpha=\{1/2,3/2\}$. 
They belong to the $SO(5)$ algebra generating the $SO(5)$ symmetry for $q=0$ and also commute separately with 
the QZE term. Note that this hidden $SU(2)\otimes SU(2)$ symmetry is related neither to the hard-core 
constraint nor to quarter filling, being rather a generic feature of $SO(5)$ fermions with QZE. 
This symmetry might be very helpful for future numerical simulations on four-component 
fermions in magnetic fields. For large $q$ this symmetry results in the mentioned iHAF. 

Growing $q$ induces PTs between the $q=0$ phases and the iHAF. 
We have obtained the ground state for various $q$ and $g$ 
by using the Density-Matrix Renormalization Group (DMRG) method 
of Ref.~\cite{Verstraete2004}, with $36$ sites, open boundary conditions, and matrix dimension $20$. 
The asymmetric phase diagram~(Fig.~\ref{fig:1}), discussed in details below, 
is characterized by a re-entrant dimerized phase.



\textit{Case of $g>0$.} 
Bosonization shows that the dimerized phase is robust at small $q$, which induces a finite chirality 
$\tau=\frac{1}{L} [(N_{3/2} + N_{-3/2})-(N_{1/2} + N_{-1/2})]\propto {\tilde q}\equiv qG/t^2$ (for ${\tilde q}\ll 1$). At large $q$ bosonization is not appropriate, and one must instead descend from the iHAF decreasing $q$. 
Spin-changing processes lead to an AF frustrating next-nearest neighbour exchange $J_2$ ($\sim \tilde q^{-2}$) between 
pseudo-spins-$1/2$, resembling a frustrated spin-$1/2$ $J_1-J_2$ AF chain, which presents a PT at $J_2/J_1\simeq 0.25$~\cite{Okamoto1992}. 
Hence a PT can be anticipated between the iHAF and the dimerized phase 
since when lowering $\tilde q$ and increasing $g$ the ratio $J_2/J$ increases. Fig.~\ref{fig:1} shows the curve $J_2/J=0.25$ 
obtained from perturbation theory, which is in good agreement with the results discussed below.
 
A strong-coupling analysis is however not a reliable proof of the existence of an actual PT, and numerical calculations must be performed to confirm it. In a finite chain the two ground states of the dimerized phase (degenerate in the thermodynamic limit) split into a unique ground state and an excited one separated by an energy gap exponentially small in the system size. Thus, for finite chains the lowest excited state in the dimerized phase is unique (for $0 < q < q_{cr}$). In contrast, the lowest excited state above the Heisenberg ground state ($q > q_{cr}$) for a finite size chain is a degenerate triplet. If the PT is of KT type, as in a frustrated spin-$1/2$ $J_1-J_2$ AF chain, a level crossing between the lowest excited singlet and triplet states should occur. 
Exact Lanczos diagonalization results, 
for up to $12$ sites with periodic boundary conditions and performing finite size scaling, 
confirm indeed this crossing. The extrapolated $q_{cr}$ lies 
in the region expected from the DMRG calculation of $\tau$ (grey region on the $g>0$ side of Fig.~\ref{fig:1}).
Although this certainly suggests a direct KT PT from the dimerized phase to the iHAF, for large $g>0$, our size 
limitations are too severe for small $g\ll 1$, and there we cannot exclude the existence of intermediate phases~\cite{footnote02}.
Hence, the iHAF chain undergoes a dimerization transition. For decreasing $q$ the dimerized phase of the pseudo-spin-$1/2$ chain adiabatically connects with the dimerized phase of spin-$3/2$ fermions, as they share similar properties. 


Although beyond the scope of this Letter, a rich physics is also expected 
in ladder-like and square lattices, since ${\cal O}({\tilde q}^{-2})$ terms induce frustrating diagonal ($J_{\times}>0$) and next-nearest neighbor ($J_2=0.5J_{\times}$) exchanges. On ladders an Ising PT is expected from a rung-singlet phase to a columnar-dimer state with growing frustration~\cite{Vekua2006}, while on a square lattice the N\'eel state may undergo a PT into a columnar-dimer state~\cite{Gelfand1989} possibly via a second-order PT showing deconfined quantum criticality~\cite{Senthil2004}.

\begin{figure}
\hspace*{-0.3cm}
\includegraphics[width=5.3cm]{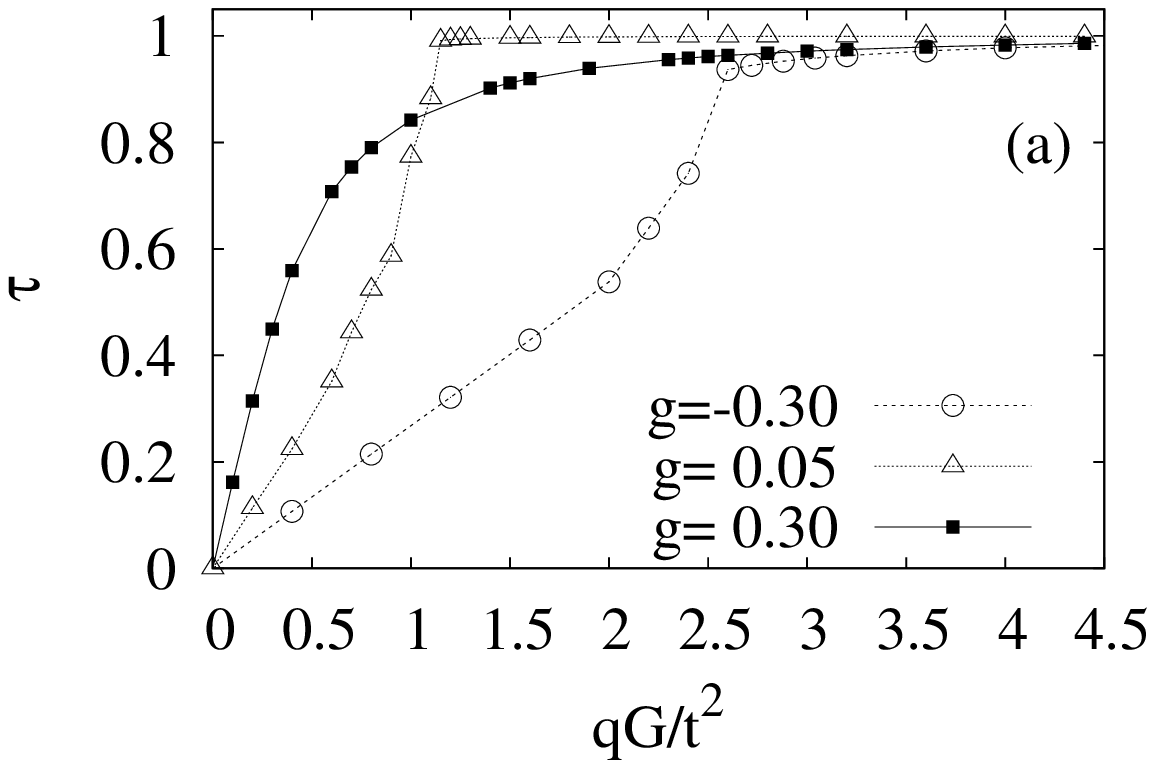} \\
\vspace*{-0.4cm}
\includegraphics[width=5.3cm]{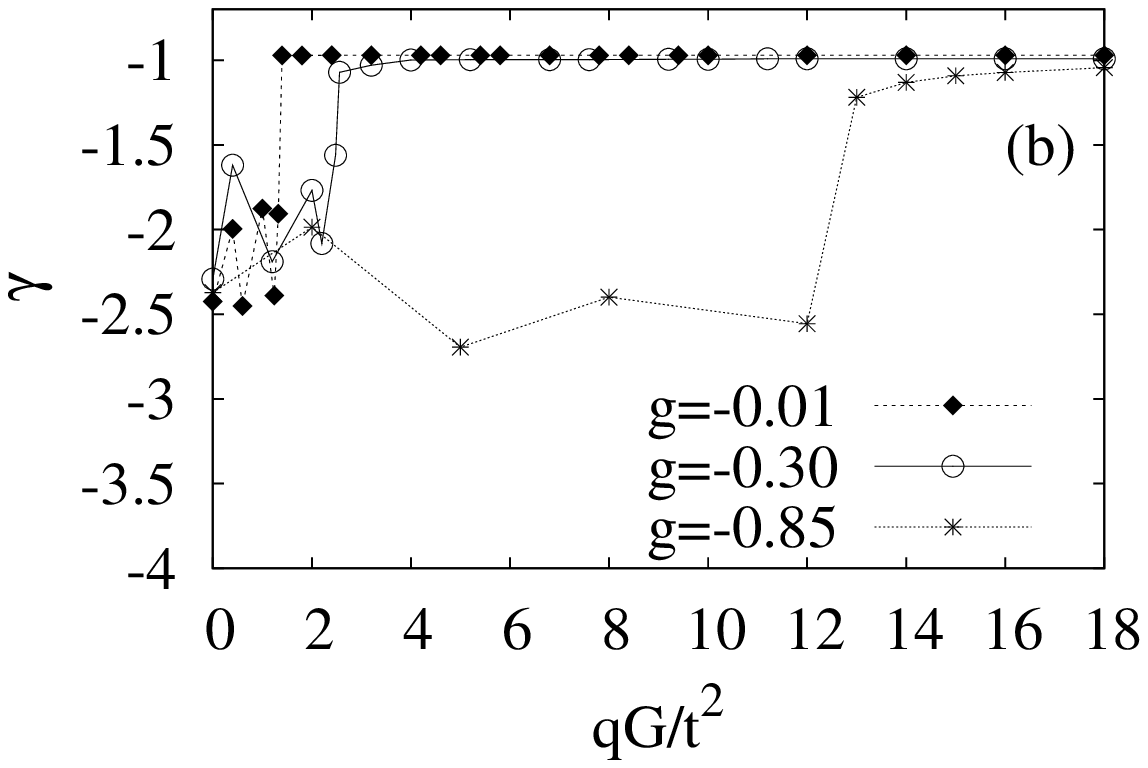}
\vspace*{-0.5cm}
\caption{Chirality $\tau$ (a) and exponent $\gamma$ (b) as a  
function of $q$, for different values of $g$. DMRG results with $36$ sites.}
\label{fig:2}
\end{figure}

We are interested in the ground-state characterization 
at finite $q$ for $g\neq 0$. The dimer phase is 
characterized by a finite $f_0$ and a singlet-triplet gap $\Delta$ 
in the magnetic excitation spectrum~\cite{Wellein1998}. Both $f_0$ and $\Delta$ 
vanish for the iHAF. We monitor as well spin-spin correlations, 
$\langle \hat S_z^i\cdot\hat S_{z}^j\rangle\propto |j-i|^{\gamma}$~\cite{footnote1}, 
and the chirality $\tau$, which ranges from $0$ ($q=0$) to $1$ (pure iHAF). 
Fig.~\ref{fig:2}(a) shows DMRG results for $\tau$. 
For weak $g>0$ the system presents still a quasi-saturation of $\tau$, which
smoothly converges to $1$ for large $g$.  
$\gamma$ behaves similarly, jumping abruptly to $-1$ for small $g>0$~(Fig.~\ref{fig:1}). 
$f_0$ and $\Delta$ evolve similarly.


\textit{Case of $g \le 0$.} In this case there is a PT 
between two gapless phases for increasing $q$.
For $g=0$ (no spin-changing collisions) the critical properties 
as a function of $q$ resemble those of a two-band model~\cite{Itakura1995}, 
where atoms with $m=\pm 1/2$ ($\pm 3/2$) act as fermions at the lowest (second) band, and  
the QZE difference $2q$ resembles the band gap.
For ${\tilde q}<{\tilde q}_0\equiv 2\ln 2\simeq 1.386$ 
the magnetic order is suppressed due to ``orbital'' effects, and the system has three massless spinons. 
On the contrary for ${\tilde q}>{\tilde q}_0$ the ``orbital'' 
degeneracy is lifted, and the manifolds $\pm 1/2$ and $\pm 3/2$ completely 
decouple in the ground state. When this occurs, $\tau$ saturates to $1$, and  
the system reduces, as mentioned above, to a pseudo-spin-$1/2$ iHAF with $J=2t^2/G$. 
Therefore at ${\tilde q}_0$ there is PT 
from a gapless spin-liquid into a gapless AF pseudo-spin-$1/2$ chain. At the phase 
transition the exponent $\gamma$ jumps to $-1$~\cite{Itakura1995}. 
We have determined at $g=0$ the critical value ${\tilde q}\simeq 1.35$, in good agreement 
with the expected value. Our results are consistent with $1-\tau \sim \sqrt{{\tilde q}_0-{\tilde q}}$ 
when approaching the PT for growing $q$. Hence at $g=0$ and ${\tilde q}_0$ there is 
a C-IC PT~\cite{Giamarchi2003} between the two gapless phases.
Moreover, $g=0$ and ${\tilde q}_0$ is a multicritical point for three phases: spin-liquid, pseudo-spin-$1/2$ iHAF, and the dimerized phase. 

The region $g<0$ smoothly connects with $g=0$ since perturbations from $g=0$ into $g<0$ are (marginally) irrelevant in the renormalization group (RG) sense~\cite{Wu2005} and the symmetry dynamically enlarges to $SU(4)$. One could hence expect that the $g<0$ region behaves similarly to the $g=0$ case for growing $q$. 
There is, however, an important distinction, since for $g\neq 0$ $\tau$ is not a good quantum number, never saturating for finite $q$. The situation resembles that of electrons in an external magnetic field in the presence of spin-non-conserving processes, which are irrelevant in the effective description but explicitly break the microscopic symmetries of the lattice model spoiling magnetization as a good quantum number~\cite{Giamarchi1988}. 
A plausible scenario for $g\le 0$ is that 
the QZE induces a C-IC PT, so that chirality-non-conserving processes 
remain irrelevant all the way, and do not modify the nature of the PT which takes place at $g=0$. 
Numerical simulations must be performed to confirm this scenario. 
The PT cannot be identified by studying $\tau$, and instead we study $\gamma$~\cite{footnote2}. 
A jump in $\gamma$, if present, will confirm the C-IC nature of the PT. 
At small $g<0$ we obtain that the PT retains the main features of that at $g=0$, 
and that indeed $\gamma$ abruptly jumps into $-1$ (Figs.~\ref{fig:1} and~\ref{fig:2}(b)). 



In summary, the QZE strongly modifies the MI phases of hard-core 
spin-$3/2$ fermions. Interestingly, contrary to bosons, a large $SU(2)\otimes SU(2)$ symmetry
remains at any $q$, resulting at large $q$ in a gapless pseudo-spin-$1/2$ iHAF.  
For $g>0$ our results suggest a KT PT from a gapped dimerized spin-$3/2$
phase into the iHAF. On the contrary, for $g\leq 0$ there is a PT 
between a gapless spin-liquid into the iHAF, which 
(at least for low $|g|$) belongs to the C-IC universality class. 
The phase diagram~(Fig.~\ref{fig:1}) has a non-trivial asymmetry characterized 
by a re-entrant dimerized phase. 
We have studied various observables, as chirality and spin correlations,  
which may reveal these phases in future experiments. 
Note that spin-changing collisions provide 
a unique opportunity to study experimentally a field-induced C-IC PT in four-component fermions, contrary 
to two-component ones where a magnetic field cannot induce a C-IC PT 
due to the conservation of magnetization.
Note also that a richer physics is expected 
for higher spins, e.g. $5/2$, where we expect for growing $q$ a sequential 
transition from a spin-$5/2$ into a pseudo-spin-$3/2$ and finally a pseudo-spin-$1/2$. 

\acknowledgements

We thank A. Kolezhuk for discussions, and the DFG for support (QUEST, 
GRK665, SFB407).


\begin{thebibliography}{99}
\bibitem{footnote-1} With a harmonic trap along the lattice, the MI phase occupy the central region. 
If the oscillator length is large-enough compared to the lattice site, surface effects are negligible and our results without trap apply.
\bibitem{footnote1} $\gamma$ is only relevant for phases with algebraically decaying spin correlations, which excludes the dimerized phase.
\end{thebibliography}
\end{document}